\documentstyle{article}

\newcommand{\be}{\begin{equation}}
\newcommand{\ee}{\end{equation}}
\newcommand{\la}{\langle}
\newcommand{\ra}{\rangle}

\def\abstract#1{\vskip 7mm 
        \begin{center}{\large Abstract}\par \smallskip
                \begin{minipage}[c]{12cm}
                        \small #1
                \end{minipage}
        \end{center}
}
\def\title#1{\begin{center}{\Large\bf #1}\end{center}}
\def\author#1{\vskip 5mm \begin{center}{#1}\end{center}}
\def\address#1{\begin{center}{\it #1}\end{center}}
\makeatletter
\@ifundefined{lesssim}{}{}
\@ifundefined{gtrsim}{}{}
\def\vereq#1#2{\lower3pt\vbox{\baselineskip1.5pt \lineskip1.5pt
\ialign{$\m@th#1\hfill##\hfil$\crcr#2\crcr\sim\crcr}}}
\makeatother

\begin{document}

\title{%
   A Status Review of Inflationary Cosmology \footnote{Invited lecture
at JGRG10, Osaka, Sept. 11 - 14, 2000, to be published in the proceedings.} 
}
\author{%
  Robert H. Brandenberger,\footnote{E-mail:rhb@het.brown.edu}
}
\address{%
  Physics Department, Brown University, \\
  Providence, R.I., USA
}

\abstract{
  The first aim of this lecture is to highlight two areas of recent
  progress in inflationary cosmology, namely reheating and the quantum
  theory of cosmological perturbations. The second aim is to discuss
  important conceptual problems for the current realizations of inflation
  based on fundamental scalar matter fields, and to present some new
  approaches at solving these problems. 
}
 
\section{Introduction}

Inflationary cosmology \cite{Guth} has become one of the cornerstones of modern cosmology. Inflation was the first theory within which it was possible to make predictions about the structure of the Universe on large scales based on causal physics. The development of the inflationary Universe scenario has opened up a new and extremely promising avenue for connecting fundamental physics with experiment.

In this lecture I wish to highlight two areas in which there have been key improvements in our understanding, namely the theory of reheating (Section 3),
and the quantum theory of cosmological perturbations (Section 4), the
cornerstone on which the recent precision comparisons between theory and observations rest. 

In spite of the remarkable success of the inflationary Universe paradigm, there are several serious conceptual problems for current models in which inflation is generated by the potential energy of a scalar matter field. These problems are discussed in Section 5.

Section 6 is a summary of some new approaches to solving the above-mentioned problems. An attempt to obtain inflation from condensates is discussed, a nonsingular Universe construction making use of higher derivative terms in the gravitational action is explained, and a framework for calculating the back-reaction of cosmological perturbations is summarized.

This short review focuses on areas of progress and on problems of inflationary cosmology. For comprehensive reviews of inflation, the reader is referred to \cite{Linde,GuthBlau,Olive,LL00}. A recent review focusing on inflationary model building in the context of supersymmetric models can be found in \cite{LR99}. An extended and more pedagogical version of these notes is \cite{RB99}.
 
\section{Basics of Inflationary Cosmology}

Most current models of inflation are based on Einstein's theory of
General Relativity with a matter source given by a scalar field $\varphi$.
Based on the cosmological principle, the metric of space-time on large distance scales can be written in Friedmann-Robertson-Walker (FRW) form:
\be
 ds^2 = dt^2 - a(t)^2 \, \left[ {dr^2\over{1-kr^2}} + r^2 (d \vartheta^2 + \sin^2 \vartheta d\varphi^2) \right] \, , 
\ee
where the constant $k$ determines the topology of the spatial sections. In the following, we shall set $k = 0$, i.e. consider a spatially flat Universe. In this case, we can without loss of generality take the scale factor $a(t)$ to be equal to $1$ at the present time $t_0$, i.e. $a(t_0) = 1$. The coordinates $r, \vartheta$ and $\varphi$ are comoving spherical coordinates.  

For a homogeneous and isotropic
Universe and setting the cosmological constant to zero, the Einstein equations  reduce to the FRW equations
\be
\left( {\dot a \over a} \right)^2 \, = \, {8 \pi G\over 3 } \rho
\ee
\be
{\ddot a\over a} \, = \, - {4 \pi G\over 3} \, (\rho + 3 p) \, ,
\ee
where $p$ and $\rho$ denote the pressure and energy density, respectively. These equations can be combined to yield the continuity equation (with Hubble
constant $H = \dot a/a$)
\be \label{cont}
\dot \rho = - 3 H (\rho + p) \, . 
\ee
The equation of state of matter is described by
a number $w$ defined by
\be
p = w \rho \, . 
\ee
 
The idea of inflation \cite{Guth} is very simple.  We assume there is a time
interval beginning at $t_i$ and ending at $t_R$ (the ``reheating time") during
which the Universe is exponentially expanding, i.e.,
\be
a (t) \sim e^{Ht}, \>\>\>\>\> t \, \epsilon \, [ t_i , \, t_R] 
\ee
with constant Hubble expansion parameter $H$.  Such a period is called  ``de
Sitter" or ``inflationary."  The success of Big Bang nucleosynthesis sets an
upper limit to the time $t_R$ of reheating, $t_R \ll t_{NS}$, 
 $t_{NS}$ being the time of nucleosynthesis.

During the
inflationary phase, the number density of any particles initially present at $t = t_i$ decays exponentially. At $t = t_R$, all of the energy which is
responsible for inflation  is released (see later) as thermal energy.  This is a
non-adiabatic process during which the entropy increases by a large factor.   

A period of inflation can solve the homogeneity
problem of standard cosmology, the reason being that during inflation
the physical size of the forward light cone exponentially expands and thus can easily become larger than the physical size of the past light cone at $t_{rec}$, the time of last scattering, thus explaining the near isotropy of the cosmic microwave background (CMB).  
Inflation also solves the flatness problem \cite{Kazanas,Guth}.

Most importantly, inflation provides a mechanism which in a causal way
generates the primordial perturbations required for galaxies, clusters and even
larger objects.  In inflationary Universe models, the Hubble radius
(``apparent" horizon) and the (``actual") horizon (the forward light cone)
do not coincide at late times.  Provided that the duration of inflation is sufficiently long, then all scales within our present apparent horizon were inside the
horizon since $t_i$.  Thus, it is in principle possible to have a causal
generation mechanism for perturbations \cite{Press,Mukh80,Lukash,Sato}.

As will be discussed in Section 4, the density perturbations produced during
inflation are due to quantum fluctuations in the matter and gravitational
fields \cite{Mukh80,Lukash}.  The amplitude of these inhomogeneities corresponds to a temperature $T_H \sim H$,
the Hawking temperature of the de Sitter phase. This leads one to expect that
at all times
$t$ during inflation, perturbations with a fixed physical wavelength $\sim
H^{-1}$ will be produced. Subsequently, the length of the waves is stretched
with the expansion of space, and soon becomes much larger than the Hubble radius $\ell_H (t) = H^{-1} (t)$.
The phases of the inhomogeneities are random.  Thus, the inflationary Universe
scenario predicts perturbations on all scales ranging from the comoving Hubble
radius at the beginning of inflation to the corresponding quantity at the time
of reheating.  In particular, provided that inflation lasts sufficiently long, perturbations on scales of galaxies and beyond will be generated. Note, however, that it is very dangerous to interpret de Sitter Hawking radiation as thermal radiation. In fact, the equation of state of this ``radiation" is not thermal  \cite{RB83}.

In most current models of inflation, the exponential expansion is driven by the potential energy density $V(\varphi)$ of a fundamental scalar matter field $\varphi$ with standard action
\begin{eqnarray} \label{fieldlag}
S_m \, &=& \int d^4x \sqrt{-g} {\cal L}_m \, \\
{\cal L}_m(\varphi) \, &=& \, {1 \over 2} D_{\mu}\varphi D^{\mu}\varphi - V(\varphi) \, ,
\end{eqnarray}
where $D_{\mu}$ denotes the covariant derivative, and $g$ is the determinant of the metric tensor. The resulting energy-momentum tensor yields the following expressions for the energy density $\rho$ and the pressure $p$:
\begin{eqnarray}
\rho (\varphi) & = & {1\over 2} \, \dot \varphi^2 + {1\over 2} \,
a^{-2}(\nabla \varphi)^2 + V (\varphi) \label{eos1} \\
p (\varphi) & = & {1\over 2} \dot \varphi^2 - {1\over 6} a^{-2}(\nabla \varphi)^2 - V(\varphi) \, . \label{eos2}
\end{eqnarray}
It thus follows that if the scalar field is homogeneous and static, but the potential energy positive, then the equation of state $p = - \rho$ necessary for exponential inflation results (see (\ref{cont})).  

Most of the current realizations of potential-driven inflation are based on satisfying the conditions 
\be \label{srcond}
\dot \varphi^2, a^{-2} (\nabla \varphi)^2 \ll V (\varphi)\, , 
\ee
via the idea of slow rolling \cite{Linde82,AS82}. Consider the equation of motion of the scalar field $\varphi$:
\be \label{eom}
\ddot \varphi + 3 H \dot \varphi - a^{-2} \bigtriangledown^2 \varphi = -
V^\prime (\varphi)\, .  
\ee
If the scalar field starts out almost homogeneous and at rest, if the Hubble damping term (the second term on the l.h.s. of (\ref{eom}) is large, and if the potential is quite flat (so that the term on the r.h.s. of (\ref{eom}) is small), then ${\dot \varphi}^2$ may remain small compared to $V(\varphi)$, in which case exponential inflation will result. Note that if the spatial gradient terms are initially negligible, they will remain negligible since they redshift.
 
Chaotic inflation \cite{Linde83} is a prototypical inflationary scenario. Consider a scalar field $\varphi$ which is very weakly coupled to itself and other fields. In this case, $\varphi$ need not be in thermal equilibrium at the Planck time, and most of the phase space for $\varphi$ will correspond to large values of $|\varphi|$ (typically $|\varphi| \gg m_{pl}$). Consider now a region in space where at the initial time $\varphi (x)$
is very large, and approximately homogeneous and static.  In this case, the energy-momentum tensor will be
immediately dominated by the large potential energy term and induce an
equation of state $p \simeq - \rho$ which leads to inflation.  Due to the
large Hubble damping term in the scalar field equation of motion, $\varphi
(x)$ will only roll very slowly towards $\varphi = 0$ (we are making the assumption that $V(\varphi)$ has a global minimum at a finite value of $\varphi$ which can then be chosen to be $\varphi = 0$).  The
kinetic energy contribution to $\rho$ and $p$ will remain small, the spatial
gradient contribution will be exponentially suppressed due to the expansion of
the Universe, and thus inflation persists.  Note that the precise form of
$V(\varphi)$ is irrelevant to the mechanism.   

It is difficult to realize chaotic inflation in conventional supergravity models since gravitational corrections to the potential of scalar fields typically render the potential steep for values of $\vert \varphi \vert$ of the order of $m_{pl}$ and larger. This prevents the slow rolling condition (\ref{srcond}) from being realizable. Even if this condition can be satisfied, there are constraints from the amplitude of produced density fluctuations which are much harder to satisfy (see Section 5).  

Hybrid inflation \cite{hybrid} is a solution to the above-mentioned problem of chaotic inflation. Hybrid inflation requires at least two scalar fields to play an important role in the dynamics of the Universe. As a toy model, consider the potential of a theory with two scalar fields $\varphi$ and $\psi$:
\be
V(\varphi, \psi) \, = \, {1 \over 4} \lambda (M^2 - \psi^2)^2 + {1 \over 2} m^2 \varphi^2 + {1 \over 2} \lambda^{'} \psi^2 \varphi^2 \, .
\ee

For values of $\vert \varphi \vert$ larger than $\varphi_c$
\be
\varphi_c \, = \, \bigl({{\lambda} \over {\lambda^{'}}} M^2 \bigr)^{1/2} \, 
\ee
the minimum of $\psi$ is $\psi = 0$, whereas for smaller values of $\varphi$ the symmetry $\psi \rightarrow - \psi$ is broken and the ground state value of $\vert \psi \vert$ tends to $M$. The idea of hybrid inflation is that $\varphi$ is slowly rolling like the inflaton field in chaotic inflation, but that the energy density of the Universe is dominated by $\psi$. Inflation terminates once $\vert \varphi \vert$ drops below the critical value $\varphi_c$, at which point $\psi$ starts to move. 

Note that in hybrid inflation $\varphi_c$ can be much smaller than $m_{pl}$ and hence inflation without super-Planck scale values of the fields is possible. It is possible to implement hybrid inflation in the context of supergravity (see e.g. \cite{hybridSG}).
 
At the present time there are many realizations of potential-driven inflation, but there is no canonical theory. A lot of attention is being devoted to implementing inflation in the context of unified theories, the prime candidate being superstring theory or M-theory. String theory or M-theory live in 10 or 11 space-time dimensions, respectively. When compactified to 4 space-time dimensions, there exist many {\it moduli} fields, scalar fields which describe flat directions in the complicated vacuum manifold of the theory. A lot of attention is now devoted to attempts at implementing inflation using moduli fields (see e.g. \cite{Banks} and references therein). 

Recently, it has been suggested that our space-time is a brane in a higher-dimensional space-time (see \cite{HW} for the basic construction).
Ways of obtaining inflation on the brane are also under active investigation (see e.g. \cite{braneinfl}).

It should also not be forgotten that inflation can arise from the purely gravitational sector of the theory, as in the original model of Starobinsky \cite{AS80} (see also Section 6), or that it may arise from kinetic terms in an effective action as in pre-big-bang cosmology \cite{PBB} or in k-inflation \cite{DM98}.  

Theories with (almost) exponential inflation generically predict an (almost) scale-invariant spectrum of density fluctuations, as was first realized in \cite{Press,Mukh80,Lukash,Sato} and then studied more quantitatively in \cite{Mukh81,flucts,BST}. Via the Sachs-Wolfe effect \cite{SW}, these density perturbations induce CMB anisotropies with a spectrum which is also scale-invariant on large angular scales.

The heuristic picture is as follows. If the inflationary period which lasts from $t_i$ to $t_R$ is almost exponential, then the physical effects which are independent of the small deviations from exponential expansion (an example of something which does depend on these deviations is effects connected with the remnant radiation density during inflation) are time-translation-invariant. This implies, for example, that quantum fluctuations at all times have the same strength when measured on the same physical length scale. 

If the inhomogeneities are small, they can described by linear theory, which implies that all Fourier modes $k$ evolve independently. The exponential expansion inflates the wavelength of any perturbation. Thus, the wavelength of perturbations generated early in the inflationary phase on length scales smaller than the Hubble radius soon becomes equal to the (``exits") Hubble radius (this happens at the time $t_i(k)$) and continues to increase exponentially. After inflation, the Hubble radius increases as $t$ while the physical wavelength of a fluctuation increases only as $a(t)$. Thus, eventually the wavelength will cross the Hubble radius again (it will ``enter" the Hubble radius) at time $t_f(k)$. Thus, it is possible for inflation to generate fluctuations on cosmological scales by causal physics.

Any physical process which obeys the symmetry of the inflationary phase and which generates perturbations will generate fluctuations of equal strength when measured when they cross the Hubble radius (see, however, Section 5.2):
\be
{{\delta M} \over M}(k, t_i(k)) \, = \, {\rm const}
\ee
(independent of $k$). Here, ${\delta M}(k, t)$ denotes the r.m.s. mass fluctuation on a length scale $k^{-1}$ at time $t$.

It is generally assumed that causal physics cannot affect the amplitude of fluctuations on super-Hubble scales (see, however, the comments at the end of Section 4.1). Therefore, the magnitude of ${{\delta M} \over M}$ can change only by a factor independent of $k$, and hence it follows that
\be \label{scaleinv}
{{\delta M} \over M}(k, t_f(k)) \, = \, {\rm const} \, ,
\ee
which is the definition of a scale-invariant spectrum \cite{HZ}.    
 
\section{Parametric Resonance and Reheating}

Reheating is an important stage in inflationary cosmology. It determines the state of the Universe after inflation and has consequences for baryogenesis, defect formation and other aspects of cosmology.

After slow rolling, the inflaton field begins to oscillate uniformly in space about the true vacuum state. Quantum mechanically, this corresponds to a coherent state of $k = 0$ inflaton particles. Due to interactions of the inflaton with itself and with other fields, the coherent state will decay into quanta of elementary particles. This corresponds to post-inflationary particle production.

Reheating is usually studied using simple scalar field toy models. The one we will adopt here consists of two real scalar fields, the inflaton $\varphi$
interacting with a massless scalar field $\chi$ representing ordinary matter. The Lagrangian is
\be
{\cal L} \, = \, {1 \over 2} \partial_\mu \varphi \partial^\mu \varphi - {1 \over 2} m^2 \varphi^2 + {1 \over 2} \partial_\mu \chi \partial^\mu \chi
- {1 \over 2} g^2 \varphi^2 \chi^2 \, , 
\ee
with $m \sim 10^{13}$GeV (see Section 4 for a justification of this choice), and $g^2$ denoting the interaction coupling constant. The bare mass and self interactions of $\chi$ are neglected. 

In the {\it elementary theory of reheating} (see e.g. \cite{DolLin} and \cite{AFW}), the decay of the inflaton was calculated using first order perturbation theory. The decay rate $\Gamma_B$ of $\varphi$ typically turns out to be much smaller than the Hubble expansion rate at the end of inflation (see \cite{RB99} for a worked example). The decay leads to a decrease in the amplitude of $\varphi$  which can be approximated by adding an extra damping term to the equation of motion for $\varphi$:
\be
{\ddot \varphi} + 3 H {\dot \varphi} + \Gamma_B {\dot \varphi} \, = \,
- V^\prime(\varphi) \, .
\ee
From the above equation it follows that as long as $H > \Gamma_B$, particle production is negligible. During the phase of coherent oscillation of $\varphi$, the energy density and hence $H$ are decreasing. Thus, eventually $H = \Gamma_B$, and at that point reheating occurs (the remaining energy density in $\varphi$ is very quickly transferred to $\chi$ particles). However, at this
point the matter temperature is much smaller than the energy scale of
inflation (reheating is a slow process).
This would imply no GUT baryogenesis, no GUT-scale defect production, and no gravitino problems in supersymmetric models with $m_{3/2} > T_R$, where $m_{3/2}$ is the gravitino mass. As we shall see, these conclusions change radically if we adopt an improved analysis of reheating.

As was first realized in \cite{TB90}, the above analysis misses an essential point. To see this, we focus on the equation of motion for the matter field $\chi$. The equations for the Fourier modes $\chi_k$ in the presence of a coherent inflaton field oscillating with amplitude $A$, 
\be
\varphi(t) \, = \, A cos(mt) \, ,
\ee
is
\be \label{reseq}
{\ddot \chi_k} + 3H{\dot \chi_k} + (k_p^2 + m_{\chi}^2 + {1 \over 2} g^2 A^2 cos(2 m t))\chi_k \, = \, 0 ,
\ee
where $k_p = k/a$ is the time-dependent physical wavenumber, and $m_{\chi}^2 = {1 \over 2} A^2$ (for other toy models a similar equation is obtained, but with a different relationship between the mass and the coefficient of the oscillating term). 

Let us for the moment neglect the expansion of the Universe. In this case, the friction term in (\ref{reseq}) drops out and $k_p$ is time-independent, and Equation (\ref{reseq}) becomes a harmonic oscillator equation with a   periodically varying mass. In the mathematics literature, this equation is called the Mathieu equation. It is well known that there is an instability. In physics, the effect is known as {\bf parametric resonance} (see e.g. \cite{parres}). At frequencies $\omega_n$ corresponding to half integer multiples of the frequency $\omega$ of
the variation of the mass, i.e.
\be
\omega_k^2 = k_p^2 + m_{\chi}^2 \, = \, ({n \over 2} \omega)^2 \,\,\,\,\,\,\, n = 1, 2, ... ,
\ee
there are instability bands with widths $\Delta \omega_n$. For values of $\omega_k$ within the instability band, the value of $\chi_k$ increases exponentially:
\be
\chi_k \, \sim \, e^{\mu t} \,\,\,\, {\rm with} \,\,\, \mu \sim {{g^2 A^2} \over {\omega}} \, .
\ee
In models of chaotic inflation $A \sim m_{pl}$. Hence, unless $g^2$ is unnaturally small (a typical value is $g^2 \sim m / m_{pl}$), it follows that
$\mu \gg H$.
 
Since the widths of the instability bands decrease as a power of the (small) coupling constant $g^2$ with increasing $n$, for practical purposes only the lowest instability band is important. Its width is
\be
\Delta \omega_k \, \sim \, g A \, .
\ee
Note, in particular, that there is no ultraviolet divergence in computing the total energy transfer from the $\varphi$ to the $\chi$ field due to parametric resonance \cite{TB90}.

It is easy to include the effects of the expansion of the Universe (see e.g. \cite{TB90,KLS94,STB95}). The main effect is that the value of $\omega_k$ becomes time-dependent. Thus, a mode slowly enters and leaves the resonance bands. As a consequence, any mode lies in the resonance band for only a finite time.  

The rate of energy transfer is given by the phase space volume of the lowest instability band multiplied by the rate of growth of the mode function $\chi_k$. Using as an initial condition for $\chi_k$ the value $\chi_k \sim H$ given by the magnitude of the expected quantum fluctuations, we obtain
\be \label{entransf}
{\dot \rho} \, \sim \, \mu ({\omega \over 2})^2 \Delta\omega_k H e^{\mu t} \, .
\ee
Hence, the energy transfer will proceed fast on the time scale
of the expansion of the Universe. There will be explosive particle production, and the energy density in matter at the end of reheating will be approximately equal to the energy density at the end of inflation.  

The above is a summary of the main physics of the modern theory of reheating.
The actual analysis can be refined in many ways (see e.g. \cite{KLS94,STB95,KLS97}, and, in the toy model considered here, \cite{GKLS}).
First of all, it is easy to take the expansion of the Universe into account
explicitly (by means of a transformation of variables), to employ an exact solution of the background model and to reduce the mode equation for $\chi_k$ to an equation which also admits exponential instabilities.

The next improvement consists of treating the $\chi$ field quantum mechanically (keeping $\varphi$ as a classical background field). At this point, the techniques of quantum field theory in a curved background can be applied. There is no need to impose artificial classical initial conditions for $\chi_k$. Instead, we may assume that $\chi$ starts in its initial vacuum state. The Bogoliubov mode mixing technique can be used to compute the number of particles at late times.

Note that the state of $\chi$ after parametric resonance is {\bf not} a thermal state. The spectrum consists of high peaks in distinct wave bands. An important question is how this state thermalizes.
For some recent progress on this issue see \cite{therm,FK00}. Since the state after explosive particle production is not a thermal state, it is useful to follow
\cite{KLS94} and call this process ``preheating" instead of reheating.

Note that the details of the analysis of preheating are quite model-dependent. In fact \cite{KLS94,KLS97}, in most models one does not get the kind of ``narrow-band" resonance discussed here, but ``broad-band" resonance. In this case, the energy transfer is even more efficient.

Recently \cite{BKM1} it has been argued that parametric resonance may lead to resonant amplification of super-Hubble-scale cosmological perturbations. The point is that in the presence of an oscillating inflaton field, the equation of motion for the cosmological perturbations takes on a similar form to the Mathieu equation discussed above (\ref{reseq}). In some models of inflation, the first resonance band includes modes with wavelength larger than the Hubble radius, leading to the apparent amplification of super-Hubble-scale modes. Such a process does not violate causality \cite{FB99} since it is driven by the inflaton field which is coherent on super-Hubble scales at the end of inflation as a consequence of the causal dynamics of an inflationary Universe. However, careful analyses for simple single-field \cite{FB99,PE99} models demonstrated that there is no net growth of the physical amplitude of gravitational fluctuations beyond what the usual theory of cosmological perturbations (see the following section) predicts.
There is increasing evidence that this conclusion holds in general for models
with purely adiabatic perturbations \cite{AB00,MLZ}. In the case of many two field models it was shown that the perturbation mode which could undergo parametric amplification during reheating is exponentially suppressed during inflation \cite{JS99,PI99}. The criterion for models (such as the one proposed in \cite{BV}) to have exponential growth of the physical amplitude of cosmological perturbations during inflation is that there is an entropy mode which is not suppressed during inflation \cite{FB00}. The resulting exponential amplification of fluctuations renders these models incompatible with the observational constraints, even including back-reaction effects \cite{ZBS00}.

\section{Quantum Theory of Cosmological Perturbations}

On scales larger than the Hubble radius the Newtonian theory of
cosmological perturbations is inapplicable, and a general
relativistic analysis is needed.  On these scales, matter is essentially frozen
in comoving coordinates.  However, space-time fluctuations can still increase
in amplitude.

In principle, it is straightforward to work out the general relativistic theory
of linear fluctuations \cite{Lifshitz}.  We linearize the Einstein  equations
\be
G_{\mu\nu} = 8 \pi G T_{\mu\nu} 
\ee
(where $G_{\mu\nu}$ is the Einstein tensor associated with the space-time
metric $g_{\mu\nu}$, and $T_{\mu\nu}$ is the energy-momentum tensor of matter)
about an expanding FRW background $(g^{(0)}_{\mu\nu} ,\, \varphi^{(0)})$:
\begin{eqnarray}
g_{\mu\nu} (\underline{x}, t) & = & g^{(0)}_{\mu\nu} (t) + h_{\mu\nu}
(\underline{x}, t) \\
\varphi (\underline{x}, t) & = & \varphi^{(0)} (t) + \delta \varphi
(\underline{x}, t) \,  
\end{eqnarray}
and pick out the terms linear in $h_{\mu\nu}$ and $\delta \varphi$ to obtain
\be \label{linein}
\delta G_{\mu\nu} \> = \> 8 \pi G \delta T_{\mu\nu} \, . 
\ee
In the above, $h_{\mu\nu}$ is the perturbation in the metric and $\delta
\varphi$ is the fluctuation of the matter field $\varphi$.  

In practice, there are many complications which make this analysis highly
nontrivial.  The first problem is ``gauge
invariance" \cite{PressVish}.   Imagine starting
with a homogeneous FRW cosmology and introducing new coordinates which mix
$\underline{x}$ and $t$.  In terms of the new coordinates, the metric now looks
inhomogeneous.  The inhomogeneous piece of the metric, however, must be a pure
coordinate (or "gauge") artifact.  Thus, when analyzing relativistic
perturbations, care must be taken to factor out effects due to coordinate
transformations.

There are various methods of dealing with gauge artifacts.  The simplest and
most physical approach is to focus on gauge invariant variables, i.e.,
combinations of the metric and matter perturbations which are invariant under
linear coordinate transformations.

The gauge invariant theory of cosmological perturbations is in principle
straightforward, although technically rather tedious. In the following I will
summarize the main steps and refer the reader to \cite{MFB92} for the details 
and further references (see also \cite{MFB92b} for a pedagogical introduction 
and \cite{Bardeen,BKP83,KoSa84,Durrer,Lyth,Hwang,EllisBruni,Salopek} for 
other approaches).

We consider perturbations about a spatially flat Friedmann-Robertson-Walker
metric
\be
ds^2 = a^2 (\eta) (d\eta^2 - d \underline{x}^2) 
\ee
where $\eta$ is conformal time (related to cosmic time $t$ by $a(\eta)  d \eta
= dt$).  At the linear level, metric perturbations can be decomposed into 
scalar modes, vector modes and tensor modes (gravitational waves). In the 
following, we will focus on the scalar modes since they are the only ones 
which couple to energy density and pressure. A scalar metric perturbation 
(see \cite{Stewart} for a precise definition)
can be written in terms of four free functions of space and time:
\be
\delta g_{\mu\nu} = a^2 (\eta) \pmatrix{2 \phi & -B_{,i} \cr
-B_{,i} & 2 (\psi \delta_{ij} + E_{,ij}) \cr} \, . 
\ee
 
The next step is to consider infinitesimal coordinate transformations
which preserve the scalar nature of $\delta g_{\mu\nu}$, and to calculate the
induced transformations of $\phi, \psi, B$ and $E$.  Then we find invariant
combinations to linear order.  (Note that there are in general no combinations
which are invariant to all orders \cite{SteWa}.)  After some algebra, it follows
that
\begin{eqnarray}
\Phi & = & \phi + a^{-1} [(B - E^\prime) a]^\prime \\
\Psi & = & \psi - {a^\prime\over a} \, (B - E^\prime)  
\end{eqnarray}
are two invariant combinations (a prime denotes differentiation
with respect to $\eta$).

Perhaps the simplest way \cite{MFB92} to derive the equations of motion for 
gauge invariant variables is to consider the linearized
Einstein equations (\ref{linein}) and to write them out in the longitudinal 
gauge defined by $B = E = 0$, in which $\Phi = \phi$ and $\Psi = \psi$, to 
directly obtain gauge invariant equations.

For several types of matter, in particular for scalar field matter,  
$\delta T^i_j \sim \delta^i_j$ 
which implies $\Phi = \Psi$.  Hence, the scalar-type cosmological perturbations
can in this case be described by a single gauge invariant variable.  The
equation of motion takes the form \cite{BST,BK84,Lyth,RBrev,Gotz} 
\be \label{conserv}
\dot \xi = O \left({k\over{aH}} \right)^2 H \xi  
\ee
where
\be
\xi = {2\over 3} \, {H^{-1} \dot \Phi + \Phi\over{1 + w}} + \Phi \, . 
\ee

The variable $w = p/ \rho$ (with $p$ and $\rho$ background pressure and energy
density respectively) is a measure of the background equation of state.  In
particular, on scales larger than the Hubble radius, the right hand side of
(\ref{conserv}) is negligible, and hence $\xi$ is constant. The above
equation (\ref{conserv}) is only true if the system has purely adiabatic
perturbations. In the case of several matter fields, entropy perturbations
can lead to terms on the right hand side of (\ref{conserv}) which are
important even on length scales much larger than the Hubble radius (see
e.g. \cite{FB00} and references therein). This is the basic reason which
allows the amplification of super-Hubble modes during reheating in such
models as discussed at the end of the previous section.

If the equation of state of matter is constant, {\it i.e.}, $w = {\rm
const}$, then $\dot \xi = 0$ implies that the relativistic potential
is time-independent on scales larger than the Hubble radius, 
i.e. $\Phi (t) = {\rm const}$. During a transition from an
initial phase with $w = w_i$  to a phase with $w = w_f$, $\Phi$ changes. 
In many cases, a good approximation to the dynamics given by (\ref{conserv}) is
\be \label{cons3}
{\Phi\over{1 + w}}(t_i)  \, = \, {\Phi\over{1 + w}}(t_f)  \, , 
\ee

To make contact with late time matter perturbations and with the Newtonian intuition, it is useful to note that, as a consequence of the Einstein constraint equations, at Hubble radius crossing $\Phi$ is a measure of the fractional density fluctuations:
\be
\Phi (k, t_H (k) ) \sim {\delta \rho\over \rho} \, ( k , \, t_H (k) ) \, .
\ee
 
The primordial perturbations in an inflationary cosmology 
are generated by quantum fluctuations (see also \cite{Bardeen2,RB84}). 
What follows is a very brief description of the unified
analysis of the quantum generation and evolution of perturbations in an 
inflationary Universe (for a detailed review see \cite{MFB92}).
The basic point is that at the linearized level, the equations describing 
both gravitational and matter perturbations can be quantized in a consistent 
way. The use of gauge invariant variables makes the analysis both physically 
clear and computationally simple. This unified description of the
generation and evolution of cosmological fluctuations goes back to the
pioneering work of Sasaki \cite{Sasaki} and Mukhanov \cite{Mukh88} but has
only fairly recently gained widespread recognition. 

The first step of this analysis is to consider the action for the linear 
perturbations about a homogeneous and isotropic background, i.e. to 
expand the gravitational and matter action $S(g_{\mu \nu}, \varphi)$ to 
quadratic order in the fluctuation variables $h_{\mu \nu}, \delta \varphi$
\be
S(g_{\mu \nu}, \varphi) \, = \, S_0(g^{(0)}_{\mu\nu}, \varphi^{(0)}) \, + 
S_2(h_{\mu \nu}, \delta \varphi; g^{(0)}_{\mu\nu}, \varphi^{(0)}) \, ,
\ee
where $S_2$ is quadratic in the perturbation variables. Focusing on the scalar 
perturbations, it turns out that one can express the resulting $S_2$ in terms 
of a single  gauge invariant variable $v$ which is a combination of metric
and matter perturbations.
\be
v \, = \, a \bigl( \delta \varphi + {{\varphi^{(0)'}} \over {\cal{H}}}
\Phi \bigr) \, .
\ee
In the above, a prime denotes the derivative with respect to conformal time, 
and ${\cal H} \, = \, a^{\prime} / a$. It turns out that, after a lot of 
algebra, the action $S_2$ reduces to the action of a single gauge invariant 
free scalar field (namely $v$) with a time dependent mass \cite{Mukh88,Sasaki} 
(the time dependence reflects the expansion of the background space-time) 
\be
S_2 \, = \, {1 \over 2} \int dt d^3x \bigl( v^{\prime 2} - (\nabla v)^2 + {{z^{\prime \prime}} \over z} v^2 \bigr) \, ,
\ee
with
\be
z \, = \, {{a \varphi_0^{\prime}} \over {\cal H}} \, .
\ee
This result is not surprising. Based on the study of classical cosmological 
perturbations, we know that there is only one field degree of freedom for the 
scalar perturbations. Since at the linearized level there are no mode 
interactions, the action for this field must be that of a free scalar field. 
We can thus use standard methods to quantize this theory. If we employ 
canonical quantization, then the mode functions of the field operator obey the same equations as we derived in the gauge-invariant analysis of 
classical relativistic perturbations. 

The time dependence of the mass leads to 
equations which have growing modes which correspond to particle
production
or equivalently to the generation and amplification of fluctuations. 
Since inflation exponentially dilutes the density of pre-existing matter, it is reasonable to assume that the perturbations start off (e.g. at the beginning of inflation) in the vacuum state (defined as a state with no particles with respect to a local 
comoving observer). The state defined this way will not be the vacuum state 
from the point of view of an observer at a later time. The Bogoliubov mode 
mixing technique can be used to calculate the number density of particles at 
a later time. In particular, expectation values of field operators such as 
the power spectrum can be computed.

The resulting power spectrum gives the following result for the mass perturbations at time $t_i(k)$:
\be \label{inmass}
\left( {\delta M\over M} \right)^2 \, (k, t_i (k)) \sim k^3 \left({V^\prime
(\varphi_0) \delta \tilde \varphi (k, t_i (k))\over \rho_0} \right)^2 \sim
\left({V^\prime (\varphi_0) H\over \rho_0 } \right)^2 \, . 
\ee
 
If the background scalar field is rolling slowly, then
\be \label{slowroll1}
V^\prime (\varphi_0 (t_i (k))) =  3 H | \dot \varphi_0 (t_i (k)) | \, .
\ee
and
\be \label{slowroll2}
(1 + p/\rho)(t_i(k)) \, \simeq \, \rho_0^{-1} {\dot \varphi_0^2}(t_i(k)) \, .
\ee
Combining (\ref{cons3}), (\ref{inmass}), (\ref{slowroll1}) and (\ref{slowroll2}) and we get
\be
{\delta M\over M} (k, \, t_f (k))  \sim \, {3 H^2 | \dot \varphi_0
(t_i (k)) |\over{\dot \varphi^2_0 (t_i (k))}} =  {3H^2\over{| \dot \varphi_0 (t_i (k))|}} 
\ee
This result can now be evaluated for specific models of inflation to find the
conditions on the particle physics parameters which give a value
\be \label{obs2}
{\delta M\over M} (k, \, t_f (k))  \sim 10^{-5} 
\ee
which is required if quantum fluctuations from inflation are to provide the
seeds for galaxy formation and agree with the CMB anisotropy limits.

For chaotic inflation with a potential
\be
V (\varphi) = {1\over 2} m^2 \varphi^2 \, , 
\ee
we can solve the slow rolling equations for the inflaton to obtain
\be \label{massconstr}
{\delta M\over M} (k, t_f (k))  \sim 10^2 {m\over m_{pl}} 
\ee
which implies that $m \sim 10^{13} \, {\rm GeV}$ to agree with (\ref{obs2}).
Similarly, for a quartic potential  
with coupling constant $\lambda$, the condition  
$\lambda \leq 10^{-12}$ is required in order not to conflict with observations.
Thus, in both examples one needs a very small parameter in the particle physics model. It  has been shown quite generally \cite{Freese}  that
small parameters are required if inflation is to solve the fluctuation problem.

To summarize the main results of the analysis of density
fluctuations in inflationary cosmology:
\begin{enumerate}
\item{} Quantum vacuum fluctuations in the de Sitter phase of an inflationary
Universe are the source of perturbations.
\item{} As a consequence of the change in the background equation of state, the evolution outside the Hubble radius produces a large
amplification of the perturbations.  In fact, unless the particle physics
model contains very small coupling constants, the predicted fluctuations are
in excess of those allowed by the bounds on cosmic microwave anisotropies.
\item{} The quantum generation and classical evolution of fluctuations can be treated in a unified manner. The formalism is no more complicated that the study of a free scalar field in a time dependent background.
\item{} Inflationary Universe models generically produce an approximately scale invariant Harrison-Zel'dovich spectrum
\be
{\delta M\over M} (k , t_f (k) ) \, \simeq \, {\rm const.} 
\ee
\end{enumerate}
  
\section{Problems of Inflationary Cosmology}

\subsection{Fluctuation Problem}

A generic problem for all realizations of potential-driven inflation studied up to now concerns the amplitude of the density perturbations which are induced by quantum fluctuations during the period of exponential expansion \cite{flucts,BST}. From the amplitude of CMB anisotropies measured by COBE, and from the present amplitude of density inhomogeneities on scales of clusters of galaxies, it follows that the amplitude of the mass fluctuations ${\delta M} / M$ on a length scale given by the comoving wavenumber $k$ at the time $t_f(k)$ when that scale crosses the Hubble radius in the FRW period is of the order $10^{-5}$. 

However, as was discussed in detail in the previous section, the present realizations of inflation based on scalar quantum field matter generically \cite{Freese} predict a much larger value of these fluctuations, unless a parameter in the scalar field potential takes on a very small value. For example, as discussed at the end of the previous section, in a single field chaotic inflationary model with quartic potential
the mass fluctuations generated are of the order
$10^2 \lambda^{1/2}$. Thus, in order not to conflict with observations, a value of $\lambda$ smaller than $10^{-12}$ is required. There have been many attempts to justify such small parameters based on specific particle physics models, but no single convincing model has emerged.

With the recent discovery \cite{BKM1,FB99} that long wavelength gravitational
fluctuations may be amplified exponentially during reheating, a new aspect of
the fluctuation problem has emerged. All models in which such amplification
occurs (see e.g. \cite{FB00} for a discussion of the required criteria) are ruled out because the amplitude of the fluctuations after back-reaction has
set in is too large, independent of the value of the coupling constant \cite{ZBS00}.

\subsection{Trans-Planckian Problem}

In many models of inflation, in particular in chaotic inflation, the period of inflation is so long that comoving scales of cosmological interest today corresponded to a physical wavelength much smaller than the Planck length at the beginning of inflation. In extrapolating the evolution of cosmological perturbations according to linear theory to these very early times, we are implicitly making the assumptions that the theory remains perturbative to arbitrarily high energies and that it can be described by classical general
relativity. Both of these assumptions break down on super-Planck scales.
Thus the question arises as to whether the predictions of the theory are
robust against modifications of the model on super-Planck scales.

A similar problem occurs in black hole physics \cite{Jacobson}. The mixing between the modes falling towards the black hole from past infinity and the Hawking radiation modes emanating to future infinity takes place in the extreme ultraviolet region, and could be sensitive to super-Planck physics. However, in the case of black holes it has been shown that for sub-Planck wavelengths at
future infinity, the predictions do not change under a class of drastic
modifications of the physics described by changes in the dispersion relation
of a free field \cite{Unruh95,CJ}. 
  
As was recently \cite{BM,MB} discovered, the result in the case of inflationary
cosmology is different: the spectrum of fluctuations may depend quite sensitively on the dispersion relation on super-Planck scales. If we take
the initial state of the fluctuations at the beginning of inflation to be given
by the state which minimizes the energy density, then for certain of the dispersion relations considered in \cite{CJ}, the spectrum of fluctuations
changes quite radically. The index of the spectrum can change (i.e. the
spectrum is no longer scale-invariant) and oscillations in the spectrum may
be induced. Note that for the class of dispersion relations considered in \cite{Unruh95} the predictions are the standard ones. 

The above results may be bad news for people who would like to consider
scalar-field driven inflationary models as the ultimate theory. However, the
positive interpretation of the results is that the spectrum of fluctuations may provide a window on super-Planck-scale physics. The present observations can
already be interpreted in the sense \cite{BM2} that the dispersion relation of the effective field theory which emerges from string theory cannot differ
too drastically from the dispersion relation of a free scalar field. 

\subsection{Singularity Problem}

Scalar field-driven inflation does not eliminate singularities from cosmology. Although the standard assumptions of the Penrose-Hawking theorems break down if matter has an equation of state with negative pressure, as is the case during inflation, nevertheless it can be shown that an initial singularity persists in inflationary cosmology \cite{Borde}. This implies that the theory is incomplete. In particular, the physical initial value problem is not defined.
 
\subsection{Cosmological Constant Problem}

Since the cosmological constant acts as an effective energy density, its value is bounded from above by the present energy density of the Universe. In Planck units, the constraint on the effective cosmological constant $\Lambda_{eff}$ is
(see e.g. \cite{cosmorev})
\be
{{\Lambda_{eff}} \over {m_{pl}^4}} \, \le \, 10^{- 122} \, .
\ee
This constraint applies both to the bare cosmological constant and to any matter contribution which acts as an effective cosmological constant.

The true vacuum value (taken on to be specific at $\varphi = 0$) of the potential $V(\varphi)$ acts as an effective cosmological constant. Its value is not constrained by any particle physics requirements (in the absence of special symmetries). Thus there must be some as yet
unknown mechanism which cancels (or at least almost completely cancels) 
the gravitational effects of any vacuum potential energy of any scalar field. However, scalar field-driven inflation relies on the almost constant potential
energy $V(\varphi)$ during the slow-rolling period acting gravitationally. How can one be sure that the unknown mechanism which cancels the gravitational effects of $V(0)$ does not also cancel the gravitational effects of $V(\varphi)$ during the slow-rolling phase? This problem is the Achilles heel of any
scalar field-driven inflationary model.

Supersymmetry alone cannot provide a resolution of this problem. It is true
that unbroken supersymmetry forces $V(\varphi) = 0$ in the supersymmetric vacuum. However, supersymmetry breaking will induce a non-vanishing $V(\varphi)$ in the true vacuum after supersymmetry breaking, and a cosmological constant problem of at least 60 orders of magnitude remains even with the lowest scale
of supersymmetry breaking compatible with experiments.

\section{New Avenues}

In the light of the problems of potential-driven inflation discussed in the previous sections, it is of utmost importance to study realizations of
inflation which do not require fundamental scalar fields, or new avenues towards early Universe cosmology altogether which, while maintaining (some of) the successes of inflation, address and resolve some of its difficulties. 

Two examples of new avenues to early Universe cosmology which do not involve
conventional inflation are the pre-big-bang scenario \cite{PBB}, and the
varying speed of light postulate \cite{Moffatt,AM99}. The pre-big-bang scenario is a model in which the Universe starts in an empty and flat dilaton-dominated phase which leads to pole-law inflation. A nice feature of this theory is that the mechanism of super-inflationary expansion is completely independent of a potential and thus independent of the cosmological constant issue. The scenario, however, is confronted with a graceful exit problem \cite{PBBEP}, and the initial conditions need to be very special \cite{PBBIC} (see, however, the
discussion in \cite{BDV}). 

It is also easy to realize that a theory in which the speed of light is much larger in the early Universe than at the present time can lead to a solution of the horizon and flatness problems of standard cosmology and thus can provide an alternative to inflation for addressing them. For realizations of this
scenario in the context of the brane world ideas see e.g. \cite{Kiritsis1,Kiritsis2,Stephon}.

String theory may lead to a natural resolution of some of the puzzles of inflationary cosmology. This is an area of active research. The reader is referred to \cite{Banks} for a review of recent studies of obtaining inflation with moduli fields, and to \cite{braneinfl} for attempts to obtain inflation with branes. Below, three more conventional approaches to addressing some of the problems of inflation will be summarized.
 
\subsection{Inflation from Condensates}

At the present time there is no direct observational evidence for the existence of fundamental scalar fields in nature (in spite of the fact that most attractive unified theories of nature require the existence of scalar fields in the low energy effective Lagrangian). Scalar fields were initially introduced to particle physics to yield an order parameter for the symmetry breaking phase transition. Many phase transitions exist in nature; however, in all cases, the order parameter is a condensate. Hence, it is useful to consider the possibility of obtaining inflation using condensates, and in particular to ask if this would yield a different inflationary scenario.

The analysis of a theory with condensates is intrinsically non-perturbative. The expectation value of the Hamiltonian $\la H \ra$ of the theory contains terms with arbitrarily high powers of the expectation value $\la \varphi \ra$ of the condensate. A recent study of the possibility of obtaining inflation in a theory with condensates was undertaken in \cite{BZ97} (see also \cite{BM92} for some earlier work). Instead of truncating the expansion of $\la H \ra$ at some arbitrary order, the assumption was made that the expansion of $\la H \ra$ in powers of $\la \varphi \ra$ is asymptotic and, specifically, Borel summable (on general grounds one expects that the expansion will be asymptotic - see e.g. \cite{ARZ})
\begin{eqnarray} \label{BZpot}
\la H\ra \, &=& \, \sum_{n=0}^{\infty} {n!} (-1)^n a_n \la \varphi^n \ra \\
&=& \, \int_0^{\infty} ds {{f(s)} \over {s(sm_{pl} + \la \varphi \ra)}} e^{-1/s} \, .
\end{eqnarray}

The cosmological scenario is as follows: the expectation value $\la \varphi \ra$ vanishes at times before the phase transition when the condensate forms. Afterwards, $\la \varphi \ra$ evolves according to the classical equations of motion with the potential given by (\ref{BZpot}) (assuming that the kinetic term   assumes its standard form).
It can easily be checked that the slow rolling conditions are satisfied. However, the slow roll conditions remain satisfied for all values of $\la \varphi \ra$, thus leading to a graceful exit problem - inflation will never terminate.

However, correlation functions, in particular $\la \phi^2 \ra$, are in general infrared divergent in the de Sitter phase of an expanding Universe. One must therefore introduce a phenomenological cutoff parameter $\epsilon(t)$ into the vacuum expectation value (VEV), and replace $\la \varphi \ra$ by $\la \varphi \ra \, / \, \epsilon$. It is natural to take $\epsilon(t) \sim H(t)$ (see e.g. \cite{AL82b,VilFord}). Hence, the dynamical system consists of two coupled functions of time $\la \varphi \ra$ and $\epsilon$. A careful analysis shows that a graceful exit from inflation occurs precisely if $\la H \ra$ tends to zero when $\la \varphi \ra$ tends to large values. 

As is evident, the scenario for inflation in this composite field model is very different from the standard potential-driven inflationary scenario. It is particularly interesting that the graceful exit problem from inflation is linked to the cosmological constant problem. Note that models of inflation based
on composites presumably do not suffer from the trans-Planckian problem,
the reason being that the effective field theory which describes the
strongly interacting system is time-translation-invariant during the de Sitter phase. The physical picture is that mode interactions are essential, and are
responsible for generating the fluctuations on a scale $k$ when this scale
leaves the Hubble radius at time $t_i(k)$.
 
\subsection{Nonsingular Universe Construction}

Another possibility of obtaining inflation is by making use of
a modified gravity sector. More specifically, we can add to the
usual gravitational action higher derivative curvature terms. These
terms become important only at high curvatures. As realized a long
time ago \cite{AS80}, specific choices of the higher derivative terms
can lead to inflation. It is well motivated to consider effective
gravitational actions with higher derivative terms when studying
the properties of space-time at large curvatures, since
any effective action for classical gravity obtained
from string theory, quantum gravity, or by integrating out matter
fields, contains such terms. In our context, the interesting question
is whether one can obtain a version of inflation avoiding some of the
problems discussed in the previous section, specifically whether one
can obtain nonsingular cosmological models.  

Most higher derivative gravity theories have much worse singularity problems than Einstein's theory. However, it is not unreasonable to expect that in the fundamental theory of nature, be it string theory or some other theory, the curvature of space-time is limited. In Ref. \cite{Markov} the hypothesis was made that when the limiting curvature is reached, the geometry must approach that of a maximally symmetric space-time, namely de Sitter space. The question now becomes whether it is possible to find a class of higher derivative effective actions for gravity which have the property that at large curvatures the solutions approach de Sitter space. A {\it nonsingular Universe construction} which achieves this goal was proposed in Refs. \cite{MB92,BMS93}. It is based on adding to the Einstein action a particular combination of quadratic invariants of the Riemann tensor chosen such that the invariant vanishes only in de Sitter space-times. This invariant is coupled to the Einstein action via a Lagrange multiplier field in a way that the Lagrange multiplier constraint equation forces the invariant to zero at high curvatures. Thus, the metric becomes de Sitter and hence explicitly nonsingular.
  
If successful, the above construction will have some very appealing 
consequences.  Consider, for example, a collapsing spatially 
homogeneous Universe.  According to Einstein's theory, this Universe 
will collapse in a finite proper time to a final ``big crunch" singularity.
In the new theory, however, the Universe will approach a de Sitter model as 
the curvature increases. If the 
Universe is closed, there will be a de Sitter bounce followed by 
re-expansion.  Similarly, spherically 
symmetric vacuum solutions of the new equations of motion will presumably be nonsingular, i.e., black holes 
would have no singularities in their centers. In two dimensions, this construction has been successfully realized \cite{TMB93}.

The {\it nonsingular Universe construction} of \cite{MB92,BMS93} and its applications to dilaton cosmology \cite{BEM98,EB99} are reviewed in a recent proceedings article \cite{BM99}. What follows is just a very
brief summary of the points relevant to the problems listed in Section 5.    

The procedure for obtaining a nonsingular Universe theory \cite{MB92} is based 
on a Lagrange multiplier construction.  
Starting from the Einstein action, one can introduce Lagrange 
multipliers fields $\varphi_i$ coupled to selected curvature invariants $I_i$ and with potentials $V_i (\varphi_i)$ chosen such that at low curvature the theory reduces to Einstein's theory, whereas at high curvatures the solutions are manifestly nonsingular. The minimal requirements for a 
nonsingular theory are that {\it all} curvature invariants remain 
bounded and the space-time manifold is geodesically complete.  

It is possible to achieve this by a two-step procedure.  First, we choose a 
curvature invariant $I_1 (g_{\mu\nu})$ (e.g. $I_1 = R$) and demand that it be 
explicitly bounded by the $\varphi_1$ constraint equation.  In a second step, we demand that as $I_1 (g_{\mu\nu})$ approaches its limiting value, the metric $g_{\mu\nu}$ approach 
the de Sitter metric $g^{DS}_{\mu\nu}$, a definite nonsingular metric 
with maximal symmetry.  In this case, all curvature invariants are 
automatically bounded (they approach their de Sitter values), and the 
space-time can be extended to be geodesically complete.
The second step can be implemented by \cite{MB92} choosing a curvature 
invariant $I_2 (g_{\mu\nu})$ with the property that 
\be
I_2 (g_{\mu\nu}) = 0 \>\> \Leftrightarrow \>\> g_{\mu\nu} = 
g^{DS}_{\mu\nu} \, ,
\ee
introducing a second Lagrange multiplier field $\varphi_2$ which couples 
to $I_2$, and choosing a potential $V_2 (\varphi_2)$ which forces $I_2$ 
to zero at large $|\varphi_2|$:
\be
S = \int d^4  x \sqrt{-g} [ R + \varphi_1 I_1 + V_1 (\varphi_1) + 
\varphi_2 I_2 + V_2 (\varphi_2) ] \, , 
\ee
with asymptotic conditions  
\begin{eqnarray}
V_1 (\varphi_1) & \sim & \varphi_1 \>\> {\rm as} \> |\varphi_1|
\rightarrow \infty \\
V_2 (\varphi_2) & \sim & {\rm const} \>\> {\rm as} \> | 
\varphi_2 |  \rightarrow \infty \label{asympt3} \\
V_i (\varphi_i) & \sim & \varphi^2_i \>\> {\rm as} \> |\varphi_i | 
\rightarrow 0 \,\,\,\, i = 1, 2 \, . \label{asympt4}
\end{eqnarray}
The first constraint renders $R$ finite, the second forces $I_2$ to zero, 
and the 
third is required in order to obtain the correct low curvature limit.

The invariant  
\be \label{inv}
I_2 = (4  R_{\mu\nu} R^{\mu\nu} - R^2 + C^2)^{1/2} \, ,
\ee
singles out the de Sitter metric among all homogeneous and isotropic 
metrics (in which case adding $C^2$, the Weyl tensor square, is 
superfluous), all homogeneous and anisotropic metrics, and all 
radially symmetric metrics.

As a specific example one can consider the action \cite{MB92,BMS93}
\be \label{act2}
S = \int d^4 x \sqrt{-g} \left[ R + \varphi_1 R - (\varphi_2 + 
{3\over{\sqrt{2}}} \varphi_1) I_2^{1/2} + V_1 (\varphi_1) + V_2 
(\varphi_2) \right] 
\ee
with
\begin{eqnarray}
V_1 (\varphi_1) & = & 12 \, H^2_0 {\varphi^2_1\over{1 + \varphi_1}} \left( 1 
- {\ln (1 + \varphi_1)\over{1 + \varphi_1}} \right) \\
V_2 (\varphi_2) & =  & - 2 \sqrt{3} \, H^2_0 \, {\varphi^2_2\over{1 + 
\varphi^2_2}} \, .
\end{eqnarray}
It can be shown that all solutions of the equations of motion which follow from this action are nonsingular \cite{MB92,BMS93}. They are either periodic about Minkowski space-time $(\varphi_1, \varphi_2) = (0, 0)$ or else asymptotically approach de Sitter space ($|\varphi_2 | \rightarrow \infty$).

One of the most interesting properties of this theory is asymptotic 
freedom \cite{BMS93}, i.e., the coupling between matter and gravity goes to 
zero at high curvatures.  It is easy to add matter (e.g., dust, 
radiation or a scalar field) to the gravitational action in the standard way.
One finds that in the asymptotic de Sitter regions, the trajectories of 
the solutions projected onto the $(\varphi_1, \, \varphi_2)$ plane are 
unchanged by adding matter.  This applies, for example, in a phase of de Sitter 
contraction when the matter energy density is increasing exponentially 
but does not affect the metric.  The physical reason for asymptotic 
freedom is obvious: in the asymptotic regions of phase space, the 
space-time curvature approaches its maximal value and thus cannot be 
changed even by adding an arbitrarily high matter energy density.
Hence, there is the possibility that this theory will admit a natural suppression mechanism for cosmological fluctuations. If this were the case, then the solution of the singularity problem would at the same time help resolve the fluctuation problem of potential-driven inflationary cosmology.
 
\subsection{Back-Reaction of Cosmological Perturbations}

The linear theory of cosmological perturbations in inflationary cosmology is well studied. However, effects beyond linear order have received very little attention. Beyond linear order, perturbations can effect the background in which they propagate, an effect well known from early studies \cite{Brill} of gravitational waves. As will be summarized below, the back-reaction of cosmological perturbations in an exponentially expanding Universe acts like a negative cosmological constant, as first realized in the context of studies of gravitational waves in de Sitter space in \cite{TW}.

Gravitational back-reaction of cosmological perturbations concerns itself with the evolution of space-times which consist of small fluctuations about a symmetric Friedmann-Robertson-Walker space-time with metric $g_{\mu \nu}^{(0)}$. The goal is to study the evolution of spatial averages of observables in the perturbed space-time. In linear theory, such averaged quantities evolve like the corresponding variables in the background space-time. However, beyond linear theory perturbations have an effect on the averaged quantities. In the case of gravitational waves, this effect is well-known \cite{Brill}: gravitational waves carry energy and momentum which affects the background in which they propagate. Here, we shall focus on scalar metric perturbations.

The idea behind the analysis of gravitational back-reaction \cite{ABM1}is to expand the Einstein equations to second order in the perturbations, to assume that the first order terms satisfy the equations of motion for linearized cosmological perturbations \cite{MFB92} (hence these terms cancel), to take the spatial average of the remaining terms, and to regard the resulting equations as equations for a new homogeneous metric $g_{\mu \nu}^{(0, br)}$ which includes the effect of the perturbations to quadratic order:
\be \label{breq}
G_{\mu \nu}(g_{\alpha \beta}^{(0, br)}) \, = \, 8 \pi G \left[ T_{\mu \nu}^{(0)} + \tau_{\mu \nu} \right]\,
\ee
where the effective energy-momentum tensor $\tau_{\mu \nu}$ of gravitational back-reaction contains the terms resulting from spatial averaging of the second order metric and matter perturbations:
\be \label{efftmunu}
\tau_{\mu \nu} \, = \, < T_{\mu \nu}^{(2)} - {1 \over {8 \pi G}} G_{\mu \nu}^{(2)} > \, ,
\ee
where pointed brackets stand for spatial averaging, and the superscripts indicate the order in perturbations.

As formulated in (\ref{breq}) and (\ref{efftmunu}), the back-reaction problem is not independent of the coordinatization of space-time and hence is not well defined. It is possible to take a homogeneous and isotropic space-time, choose different coordinates, and obtain a non-vanishing $\tau_{\mu \nu}$. This ``gauge" problem is related to the fact that in the above prescription, the hypersurface over which the average is taken depends on the choice of coordinates. 

The key to resolving the gauge problem is to realize that to second order in perturbations, the background variables change. A gauge independent form of the back-reaction equation (\ref{breq}) can hence be derived \cite{ABM1} by defining background and perturbation variables $Q = Q^{(0)} + \delta Q$ which do not change under linear coordinate transformations. Here, $Q$ represents collectively both metric and matter variables. The gauge-invariant form of the back-reaction equation then looks formally identical to (\ref{breq}), except that all variables are replaced by the corresponding gauge-invariant ones. We will follow the notation of \cite{MFB92}, and use as gauge-invariant perturbation variables the Bardeen potentials \cite{Bardeen} $\phi$ and $\Psi$ which in longitudinal gauge coincide with the actual metric perturbations $\delta g_{\mu \nu}$. Calculations hence simplify greatly if we work directly in  longitudinal gauge. These calculations have been confirmed \cite{WA} by working in a completely different gauge, making use of the covariant approach.
 
In \cite{ABM2}, the effective energy-momentum tensor $\tau_{\mu \nu}$ of gravitational back-reaction was evaluated for long wavelength fluctuations in an inflationary Universe in which the matter responsible for inflation is a scalar field $\varphi$ with the potential
\be
V(\varphi) \, = \, {1 \over 2} m^2 \varphi^2 \, .
\ee
Since in this model there is no anisotropic stress, the perturbed metric in longitudinal gauge can be written \cite{MFB92} in terms of a
single gravitational potential $\phi$
\be
ds^2 =  (1+ 2 \phi) dt^2 - a(t)^2(1 - 2\phi) \delta_{i j} dx^i dx^j  \, ,
\ee
where $a(t)$ is the cosmological scale factor. 

It is now straightforward to compute $G_{\mu \nu}^{(2)}$ and 
$T_ {\mu \nu}^{(2)}$ in terms of the background fields and the metric and matter fluctuations $\phi$ and $\delta \varphi$, By taking averages and making use of (\ref{efftmunu}), the effective energy-momentum tensor $\tau_{\mu \nu}$ can be computed \cite{ABM2}.

The general expressions for the effective energy density $\rho^{(2)} = \tau^0_0$ and effective pressure $p^{(2)} = - {1 \over 3} \tau^i_i$ involve many terms. However, they greatly simplify if we consider perturbations with wavelength greater than the Hubble radius. In this case, all terms involving spatial gradients are negligible. From the theory of linear cosmological perturbations (see e.g. \cite{MFB92}) it follows that on scales larger than the Hubble radius the time derivative of $\phi$ is also negligible as long as the equation of state of the background does not change. The Einstein constraint equations relate the two perturbation variables $\phi$ and $\delta \varphi$, enabling scalar metric and matter fluctuations to be described in terms of a single gauge-invariant potential $\phi$. During the slow-rolling period of the inflationary Universe, the constraint equation takes on a very simple form and implies that $\phi$ and $\delta \varphi$ are proportional. The upshot of these considerations is that $\tau_{\mu \nu}$ is proportional to the two point function $< \phi^2 >$, with a coefficient tensor which depends on the background dynamics. In the slow-rolling approximation we obtain \cite{ABM2} 
\be
\rho^{(2)} \, \simeq \, - 4 V < \phi^2 >
\ee
and
\be
p^{(2)} \, = \, - \rho^{(2)} \, .
\ee
This demonstrates that the effective energy-momentum tensor of long-wavelength cosmological perturbations has the same form as a negative cosmological constant. 

Note that during inflation, the phase space of infrared modes is growing. Hence, the magnitude of $|\rho^{(2)}|$ is also increasing. Hence, the back-reaction mechanism may lead to a dynamical relaxation mechanism for a bare cosmological constant driving inflation \cite{RBTexas}. A similar effect holds for pure
gravity at two loop order in the presence of a bare cosmological constant \cite{TW}.

The interpretation of $\rho^{(2)}$ as a local density has been criticized, e.g. in \cite{Unruh98}. Instead of computing physical observables from a spatially averaged metric, one should compute the spatial average of physical invariants corrected to quadratic order in perturbation theory. Work on this issue
is in progress \cite{AW01} (see also \cite{AB00}).

\section{Conclusions}

Inflationary cosmology is an attractive {\it scenario}. It solves some problems of standard cosmology and leads to the possibility of a causal theory of structure formation. The specific predictions of an inflationary model of structure formation, however, depend on the specific realization of inflation, which makes the idea of inflation hard to verify or falsify. Many models of inflation have been suggested, but at the present time none are sufficiently distinguished to form a ``standard" inflationary theory.

There has been a lot of recent progress in inflationary cosmology. As explained in Section 3, a new theory of inflationary reheating (preheating) has been developed based on parametric resonance. Preheating leads to a rapid energy
transfer between the inflaton field and matter at the end of inflation, with
important cosmological consequences. A recent development in this area is
the realization that long wavelength gravitational fluctuations may be
amplified exponentially in models with an entropy perturbation mode which is
not suppressed during inflation.
 
As discussed in Section 4, there is now a consistent quantum theory of the generation and evolution of linear cosmological perturbations which describes
the origin of fluctuations from an initial vacuum state of the fluctuation modes at the beginning of inflation, and which forms the basis for the precision calculations of the power spectrum of density fluctuations and of CMB anisotropies which allow detailed comparisons with current and upcoming observations.

However, there are important conceptual problems for scalar field-driven inflationary models. Four such problems discussed in this lecture (in Section 5) are the {\it fluctuation problem}, the {\it trans-Planckian problem}, the {\it singularity problem} and the {\it cosmological constant problem}, the last of which is the Achilles heel of these inflationary models. 

It may be that a convincing realization of inflation will have to wait for an improvement in our understanding of fundamental physics. Some promising but incomplete avenues which address some of the problems mentioned above and which yield inflation exist are discussed in Section 6. 
 
\centerline{\bf Acknowledgements}

I wish to thank Profs. Jun-ichi Yokoyama and Misao Sasaki for the invitation
to present this lecture, and for their wonderful hospitality during the 
meeting. I also wish to thank the Prof. Richard MacKenzie for hospitality
at the Universit\'e de Montreal and Profs. Jim Cline and Rob Myers for
hospitality at McGill University during the time this manuscript was
completed. The author is supported in part by the US Department of Energy
under Contract DE-FG02-91ER40688, Task A.

\end{document}